\documentclass{eas}
\usepackage{graphicx}
\usepackage{amssymb}
\newcommand{\nat}{Nature}
\newcommand{\apj}{ApJ}
\newcommand{\aj}{AJ}
\newcommand{\aap}{A\&A}
\newcommand{\mnras}{MNRAS}
\begin{document}

%%-----------------------------
%%      the top matter
%%-----------------------------
\title{Multi-Wavelength Observations of Short-Duration Gamma-Ray Bursts: Recent Results} \thanks{I wish to express thanks to all my colleagues who worked diligently with me on my {\em Swift}-era afterglow papers, and the GROND team for further cooperation. Furthermore, I extend thanks to Alberto Castro-Tirado and the rest of the conference team for inviting me to Spain, as well as to the anonymous referee for a very constructive report.}
\runningtitle{Kann: Recent Results on Type I GRB Afterglows}
\author{David Alexander Kann}
\address{Max-Planck-Institut f\"ur extraterrestrische Physik, Giessenbachstra\ss e, 85748 Garching, Germany; Universe Cluster, Technische Universit\"at M\"unchen, Boltzmannstra\ss e 2, 85748 Garching, Germany}
	\email{kann@tls-tautenburg.de}
\secondaddress{Th\"uringer Landessternwarte Tautenburg, Sternwarte 5, 07778 Tautenburg, Germany}
\begin{abstract}
The number of detections as well as significantly deep non-detections of optical/NIR afterglows of Type I (compact-object-merger population) Gamma-Ray Bursts (GRBs) has become large enough that statistically meaningful samples can now be constructed. I present within some recent results on the luminosity distribution of Type I GRB afterglows in comparison to those of Type II GRBs (collapsar population), the issue of the existence of jet breaks in Type I GRB afterglows, and the discovery of {\em dark} Type I GRBs.
\end{abstract}
\maketitle
%%-----------------------------
%%      your text
%%-----------------------------
\section{Introduction}
Similar to 1997 for Type II (long-duration, collapsar population) GRBs\footnote{In this work, we follow the classification scheme detailed in Zhang \etal\ (\cite{Zhang2009}), which labels GRBs associated with the core-collapse of massive stars ``Type II'' and those which are not (but are likely associated with the mergers of compact objects) ``Type I''. This classification is independent of duration.}, the advent of the {\em Swift} satellite (Gehrels \etal\ \cite{Gehrels2004}) in the year 2005 saw the discovery of Type I GRB afterglows and their placement within the cosmological context (Gehrels \etal\ \cite{Gehrels2005}; Hjorth \etal\ \cite{Hjorth2005}; Fox \etal\ \cite{Fox2005}; Berger \etal\ \cite{Berger2005}). In the following seven years, over 50 Type I GRBs have been precisely localized by {\em Swift} BAT and XRT, and many of these have either detected optical afterglows\footnote{A few Type I GRBs have also been detected at radio wavelengths (Berger \etal\ \cite{Berger2005}, Soderberg \etal\ \cite{Soderberg2006}) and in very high-energy gamma-rays during and shortly after the prompt emission (Abdo \etal\ \cite{Abdo2010}; Ackermann \etal\ \cite{Ackermann2010}; Akerlof \etal\ \cite{Akerlof2011}; Zheng \etal\ \cite{Zheng2012}).}, or at least significantly deep limits thereon. Host-galaxy follow-up has also yielded a significant number of redshifts for these events.

In the last years, a multitude of results on Type I GRB afterglows have been published. In these proceedings, we wish to focus on three issues: The luminosity of Type I GRB afterglows, the existence and detectability of jet breaks, and {\em dark} Type I GRBs.

\section{The luminosity of Type I GRB afterglows vs. those of Type II GRBs}

\begin{figure}[!t]
\includegraphics[width=0.5\textwidth]{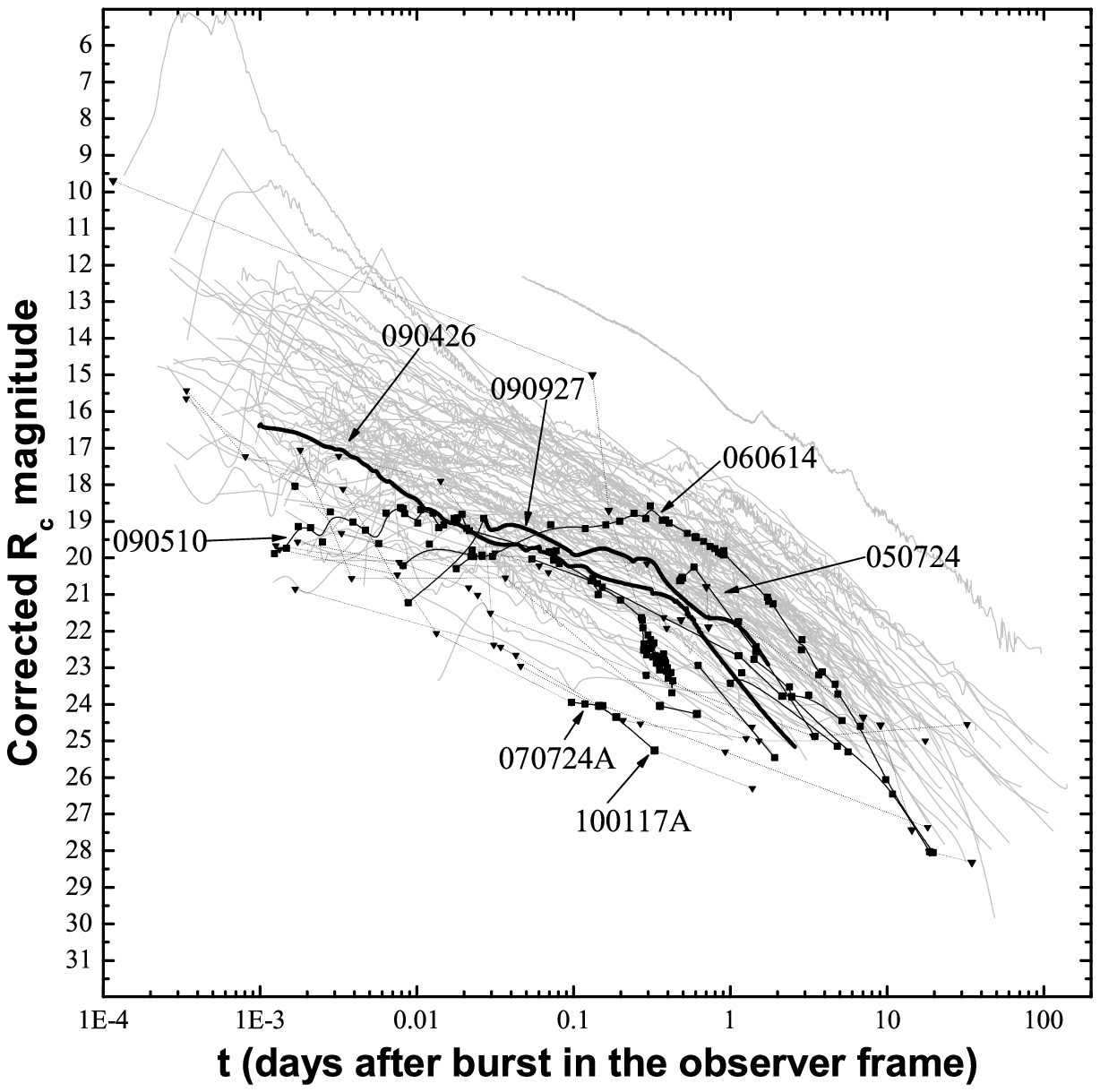}
\includegraphics[width=0.5\textwidth]{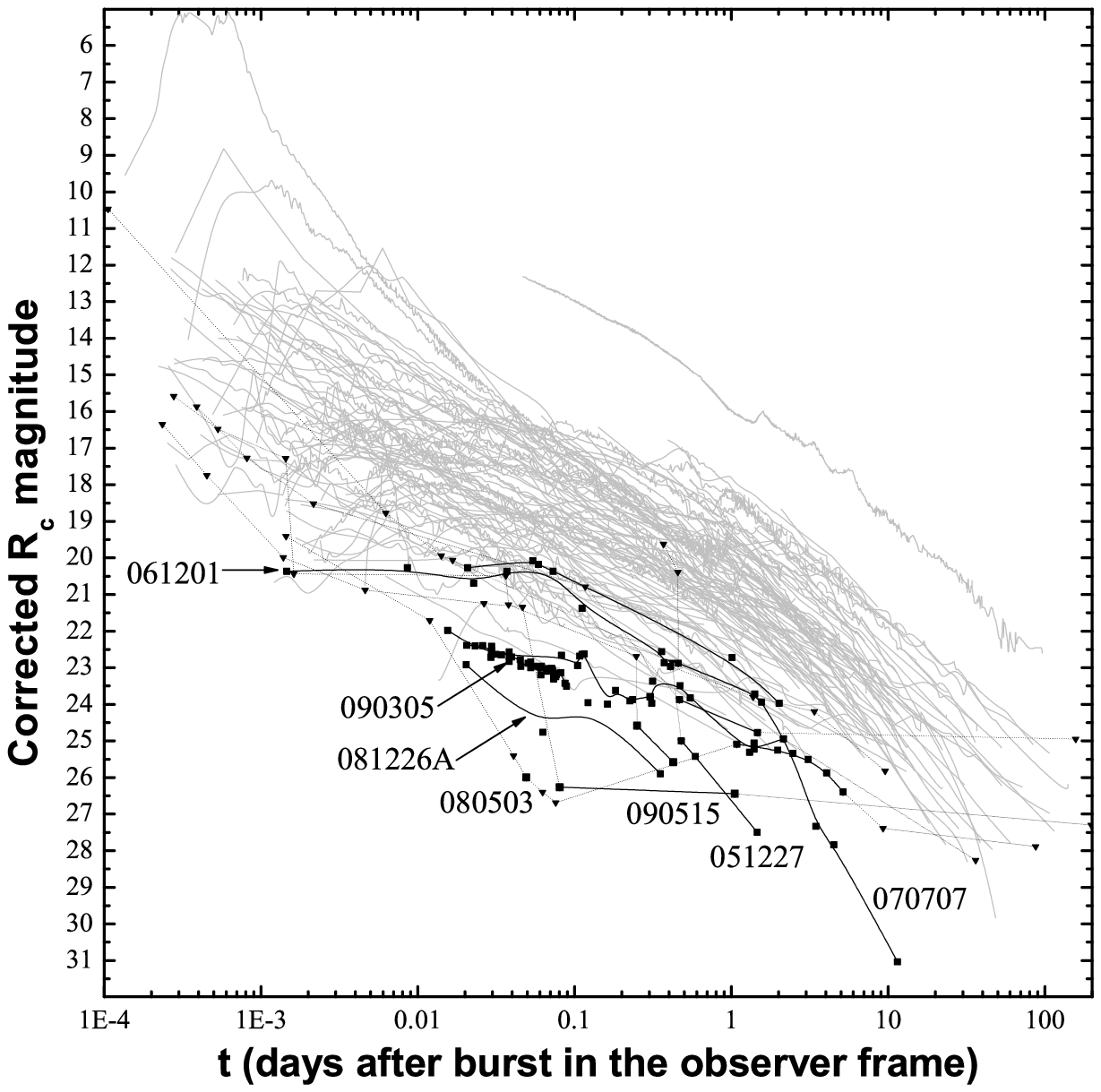}
\includegraphics[width=0.5\textwidth]{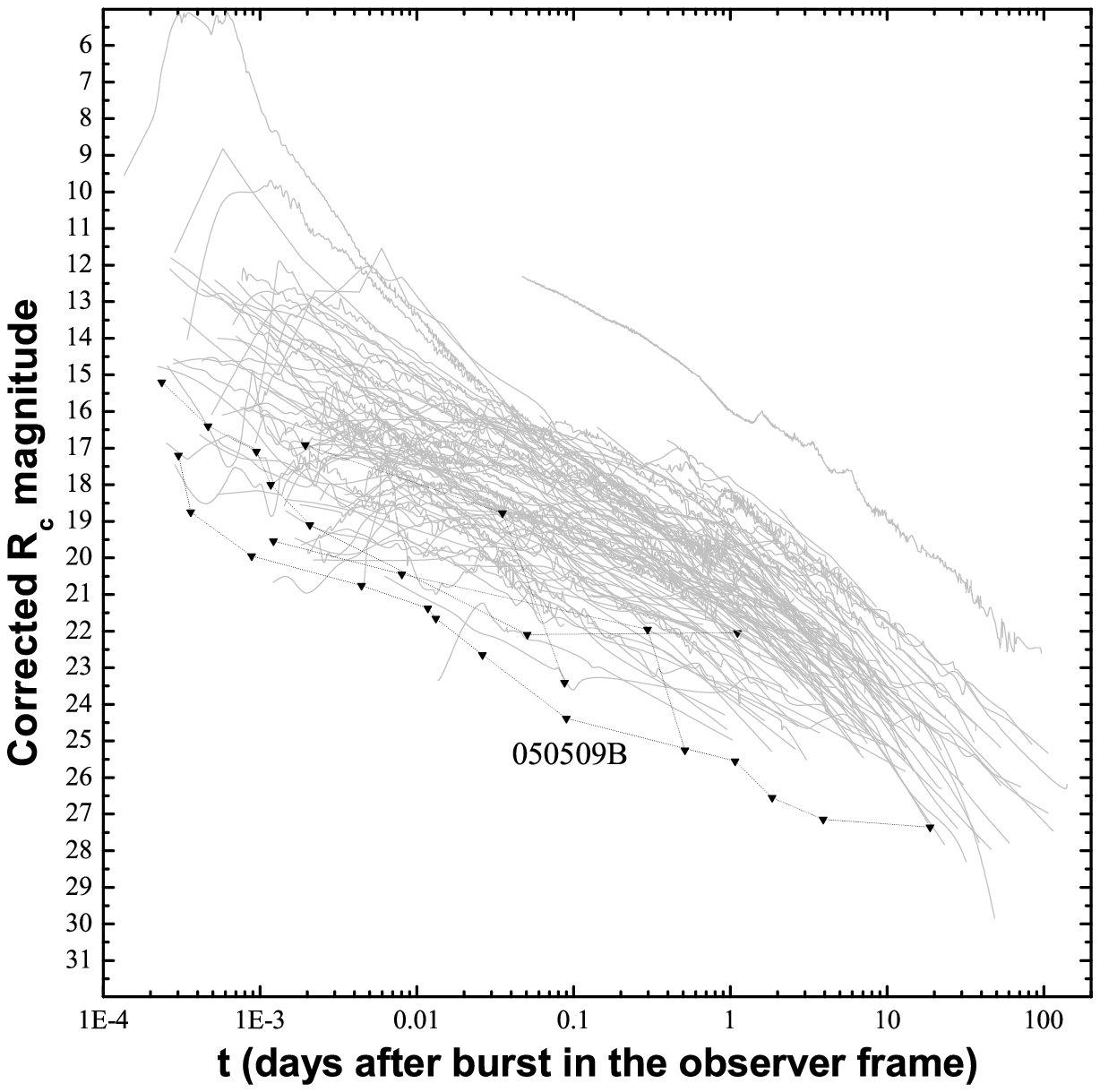}
\includegraphics[width=0.5\textwidth]{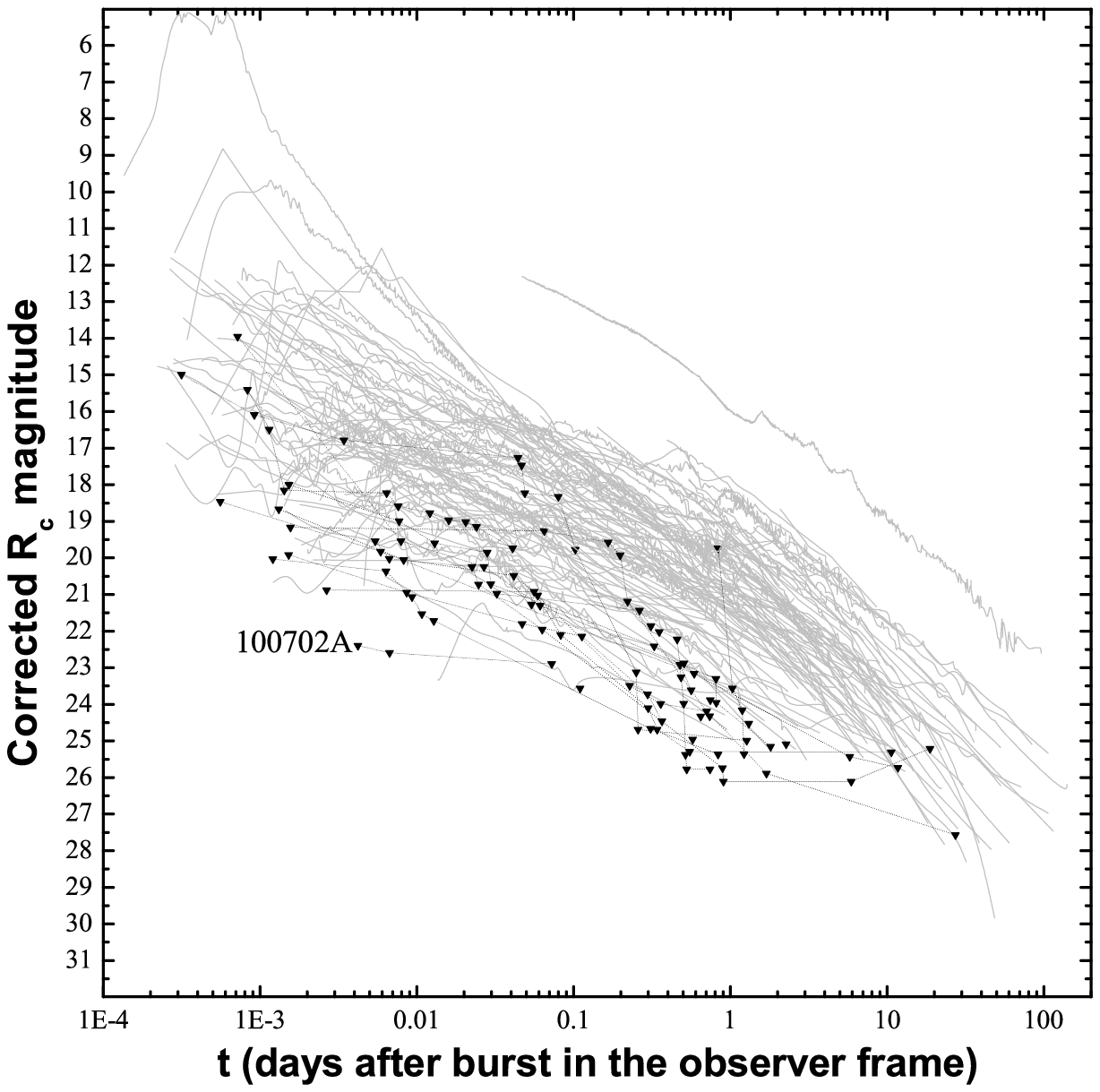}
\caption{Observed afterglows of Type I and Type II GRBs (corrected for Galactic extinction). The Type II GRB afterglow sample forms the gray ``background''. {\em Top left:} Type I GRB afterglows with detections (as well as additional upper limits for the same GRBs) and redshifts we consider secure. We additionally highlight, with thick black lines, two Type II GRB afterglows whose GRBs had very short durations, under the classical temporal dividing line. {\em Top right:} As top left, but with insecure redshifts (or simple estimates). {\em Bottom left:} Type I GRBs with upper limits only, but secure redshifts. {\em Bottom right:} As bottom left, but with insecure redshifts. Outstanding GRBs are named. From these plots, it is already clear that Type I GRB afterglows are, in the whole, fainter than those of Type II GRBs, with many having no afterglow detected to upper limits much deeper than all Type II GRB afterglow detections in our sample.}
\label{Obs}
\end{figure}

\begin{figure}[!t]
\includegraphics[width=0.5\textwidth]{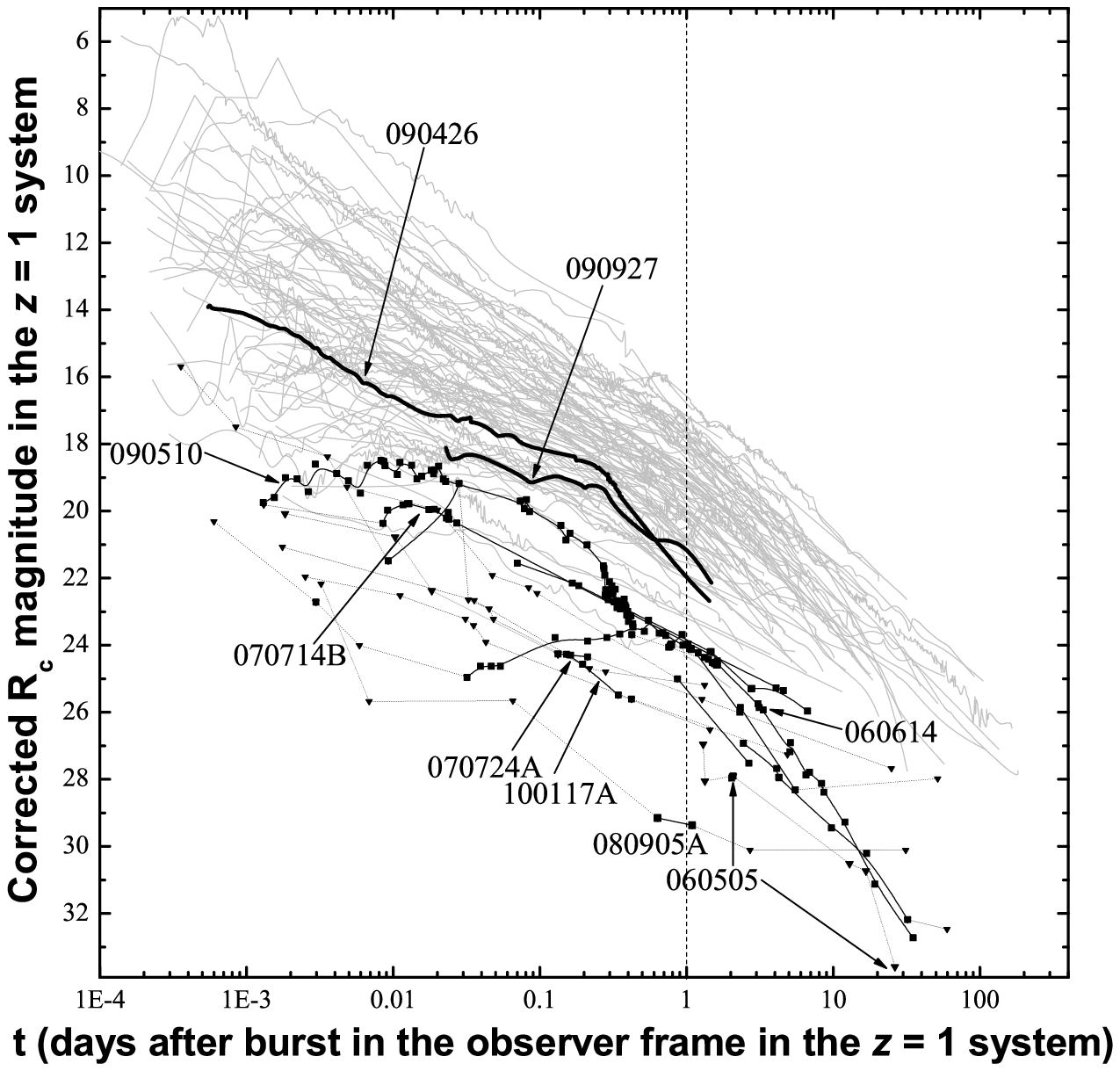}
\includegraphics[width=0.5\textwidth]{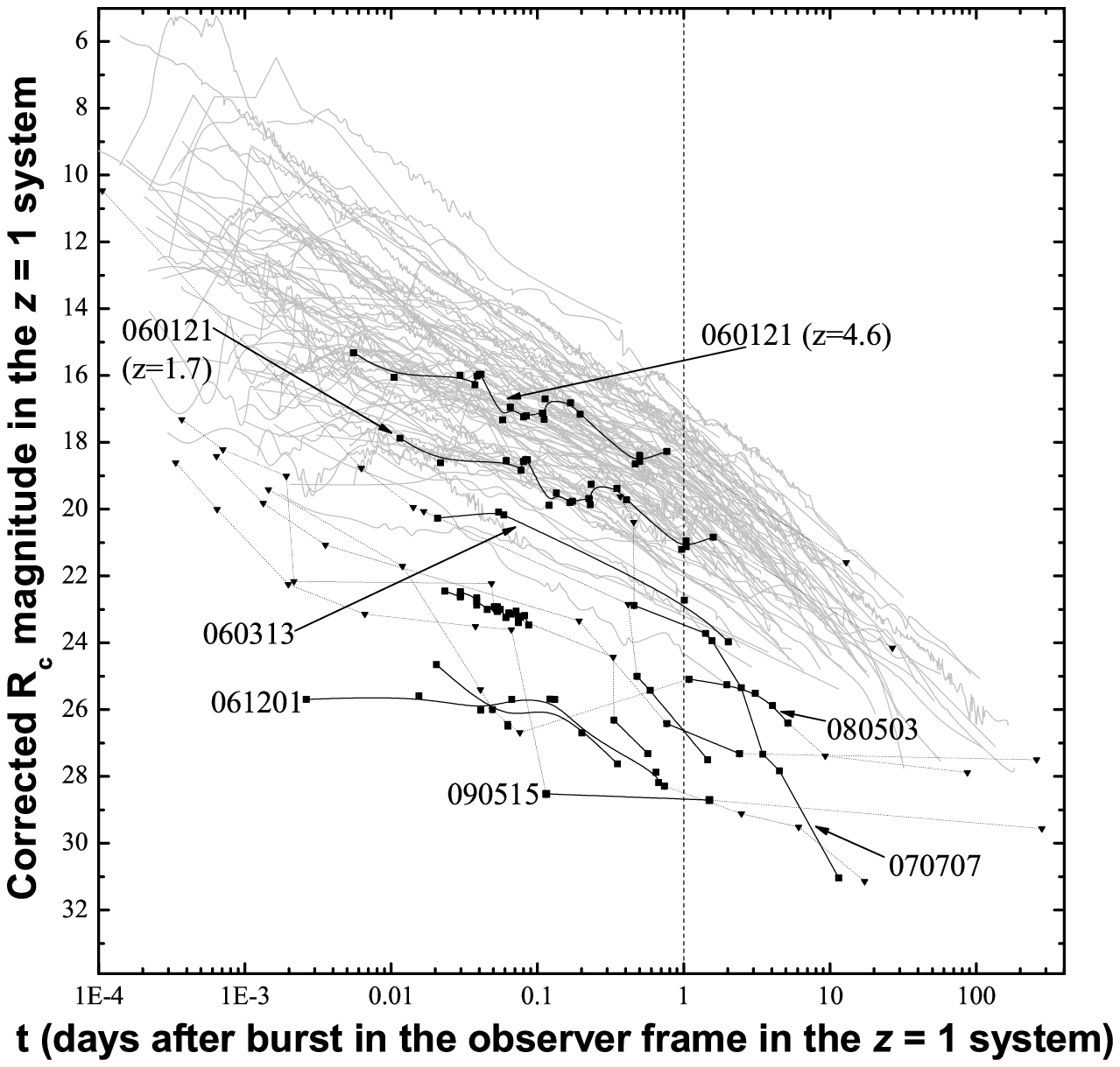}
\includegraphics[width=0.5\textwidth]{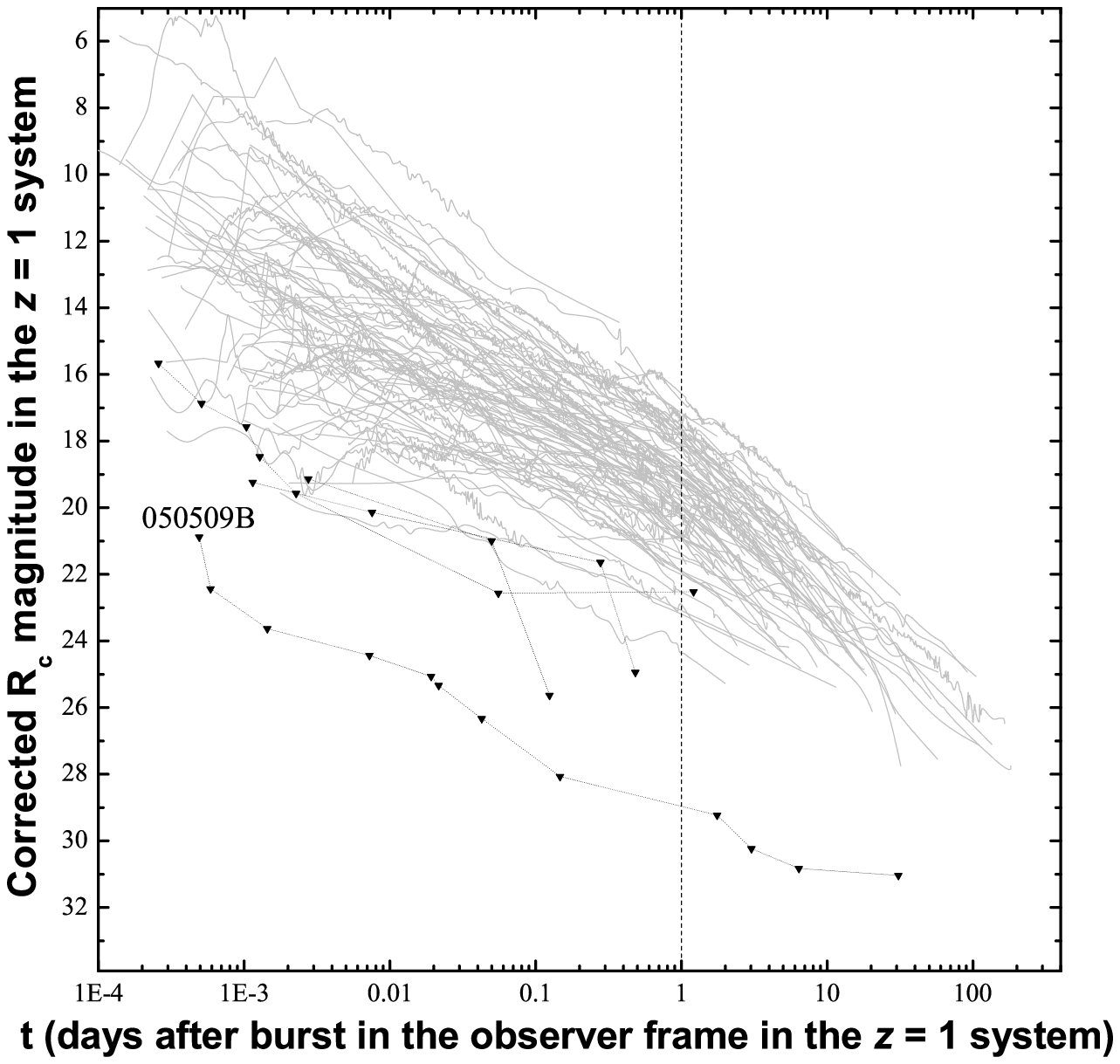}
\includegraphics[width=0.5\textwidth]{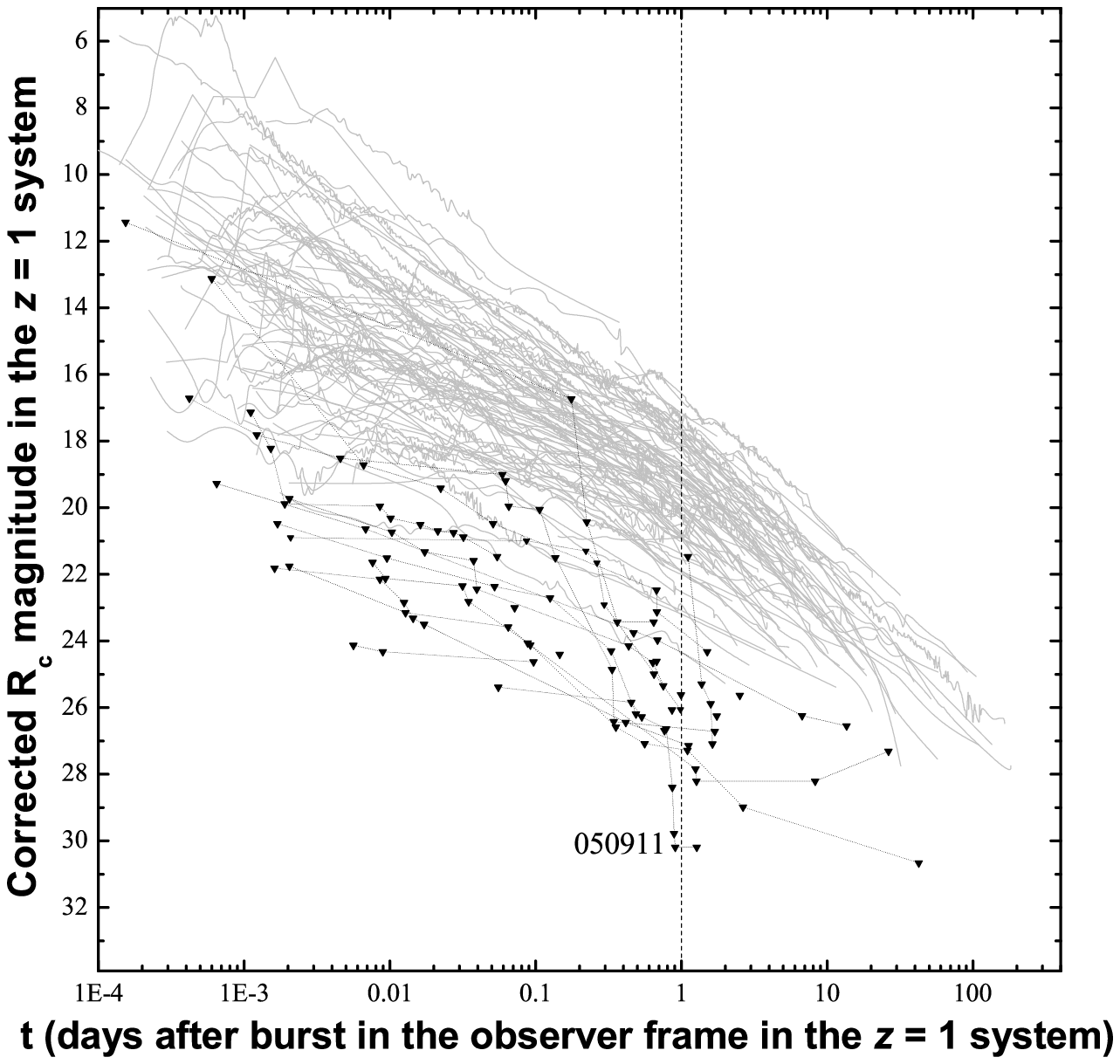}
\caption{Afterglows of Type I and Type II GRBs after correcting for rest-frame extinction (where applicable) and shifting to a common redshift of $z=1$. Sample distribution is as in Fig. \ref{Obs}. The two Type II GRB afterglows in the top left, while among the fainter ones in their class, are clearly more luminous than those of Type I GRBs, which are almost all significantly fainter than those of Type II GRBs. The only exception is GRB 060121, top right, which we propose to be a short-duration Type II GRB.}
\label{Zone}
\end{figure}

In two papers (Kann \etal\ \cite{Kann2006,Kann2010}), we studied large samples of Type II GRB afterglows,  with one aspect we focused on being the luminosity of the afterglows. Knowledge of the redshift, the rest-frame dust extinction and the intrinsic spectral slope allows a shift of the afterglow light curve to any redshift, we choose $z=1$ as a reference system to compare the afterglows. In Kann \etal\ (\cite{Kann2011}), we followed up with a study of all Type I GRB afterglows with detections or significantly deep upper limits up to the beginning of 2010, and compared these afterglows with our Type II GRB sample\footnote{See also Gehrels \etal\ \cite{Gehrels2008} and Nysewander \etal\ \cite{Nysewander2009} for similar studies of this specific issue. } (adding three more Type II GRBs in this paper to the sample). Since the publication of Kann \etal\ (\cite{Kann2011}), we have undertaken additional studies using GROND (Greiner \etal\ \cite{Greiner2008}), focusing on the post-break evolution of the short-duration Type II GRB 090426 (Nicuesa Guelbenzu \etal\ \cite{Nicuesa2011}), the extremely luminous Type I GRB 090510 (Nicuesa Guelbenzu \etal\ \cite{Nicuesa2012A}) and a large sample of Type I GRBs with GROND afterglow follow-up (Nicuesa Guelbenzu \etal\ \cite{Nicuesa2012B}).

In Figs. \ref{Obs} and \ref{Zone}, we plot our afterglow light curve samples. Hereby, the Type II GRB afterglows form a gray ``background'' which we will not discuss further. We divide our Type I GRB sample into four different subsamples, depending on whether an optical afterglow has been discovered or not, and whether we consider the redshift of the GRB secure, or if it is insecure or just estimated (see Kann \etal\ \cite{Kann2011} for more details). 

Already from Fig. \ref{Obs}, it is clear that Type I GRB afterglows are generally fainter than those of Type II GRBs. In the figure, we highlight two events, GRB 090426 and GRB 090927 (Nicuesa Guelbenzu \etal\ \cite{Nicuesa2011,Nicuesa2012B}) which had short durations, under or on the classic $2$s dividing line (Kouveliotou \etal\ \cite{Kouveliotou1993}), but which are considered to be Type II GRBs (see also Levesque \etal\ \cite{Levesque2010}; Xin \etal\ \cite{Xin2010} and Th\"one \etal\ \cite{Thoene2011} concerning GRB 090426).

In Fig. \ref{Zone}, all afterglows have been transformed to a common redshift of $z=1$, using the method of Kann \etal\ \cite{Kann2006}, and can be directly compared. Again, we separate the Type I GRB afterglow sample into the four subsamples delineated above. Panaitescu \etal\ (\cite{Panaitescu2001}) already predicted that Type I GRB afterglows should be significantly fainter, working on the hypothesis that these GRBs have compact-object-merger progenitors, which are likely to occur in a significantly less dense interstellar medium than their collapsar counterparts, and such large offsets have indeed been found (e.g., Fong \etal\ \cite{Fong2010}; Berger \cite{Berger2010}; Kann \etal\ \cite{Kann2011}). We fully confirm this result in our samples (and note that the distribution of luminosities for those samples with uncertain redshifts does not differ significantly from the samples with secure redshifts). We find that in the mean, the sample with detections and secure redshifts is $5.8\pm0.5$ mag fainter than the mean magnitude of the Type II GRB afterglow Golden Sample (see Kann \etal\ \cite{Kann2010} for definitions of the different Type II GRB afterglow samples), this makes it $\approx210^{+120}_{-80}$ times less luminous. This is roughly a factor of $5-20$ less luminous than Panaitescu \etal\ (\cite{Panaitescu2001}) had initially predicted. A comparison with their assumption shows that Panaitescu \etal\ (\cite{Panaitescu2001}) overestimated both the typical isotropic energy release of Type I GRBs, and also likely the typical external medium density. We note that the additional Type I GRBs added from Nicuesa Guelbenzu \etal\ (\cite{Nicuesa2012B}) are in full agreement with the luminosity distribution found so far, but for the most part do not have secure redshifts, and therefore do not contribute to the most meaningful comparison sample.

Fig. \ref{Zone} also clearly shows that while the afterglows of GRB 090426 and GRB 090927 are among the fainter Type II afterglows, they are more luminous than any of the Type I GRB afterglows, and therefore these GRBs likely belong to the collapsar population (for additional arguments, see links above as well as Grupe \etal\ \cite{Grupe2009}) despite their very short duration, possibly making them similar to the case of GRB 060121 (de Ugarte Postigo \etal\ \cite{deUgarte2006}; Levan \etal\ \cite{Levan2006}; Kann \etal\ \cite{Kann2011}). Similar to cases of possible Type I GRBs which are ``too long'', these examples show that likely Type II GRBs exist which are ``too short'' (see also Virgili \etal\ \cite{Virgili2011}), and therefore extended criteria beyond the simple (detector-dependent) duration are needed, as have been discussed by Zhang \etal\ (\cite{Zhang2009}).

\section{The issue of jet breaks in Type I GRB afterglows}

\begin{figure}[!t]
\includegraphics[width=\textwidth]{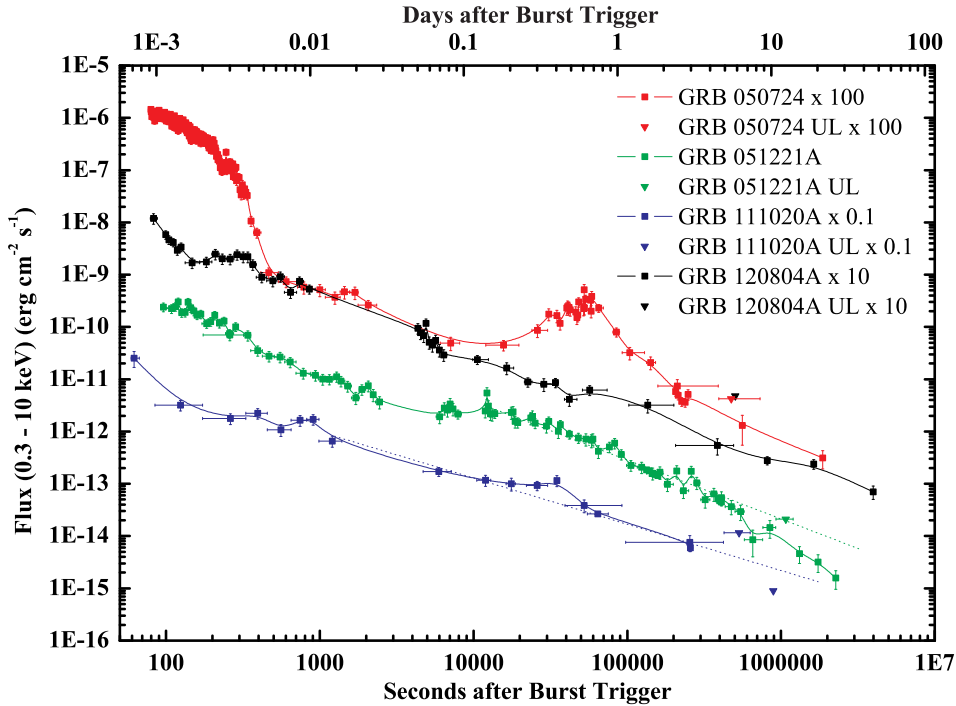}
\caption{The X-ray afterglows, as measured by {\em Swift}, {\em Chandra} and {\em XMM-Newton}, of four Type I GRBs with late-time observations ($\gtrsim1$ Ms). For clarity, the fluxes have been shifted up or down relative to the measured values by the factors given in the legend. GRBs 050724 and 120804A show unbroken decays up to several weeks after the GRB, whereas GRBs 051221A and 111020A show breaks in their light curves, which may be due to jet breaks (for GRB 111020A, the break is only found via a {\em Chandra} upper limit which is significantly deeper than the light-curve extrapolation). We highlight these breaks by extrapolating the earlier decay to late times (dotted lines). }
\label{Xray}
\end{figure}

The existence of a so-called ``jet break'' due to the collimation of the GRB emission was proposed early in the afterglow era (Rhoads \cite{Rhoads1997}), and has been studied extensively in the optical (e.g., Zeh \etal\ \cite{Zeh2006}) and the X-rays (e.g., Racusin \etal\ \cite{Racusin2009}) for Type II GRB afterglows. The question of the existence of such breaks for Type I GRB afterglows is at the same time a question of the degree of collimation such GRBs exhibit, considering the preferred model of compact-object mergers does not provide an extended envelope for the jet to propagate through which might aid in the collimation of the jet. Recent numerical studies (e.g., Rezzolla \etal\ \cite{Rezzolla2011}) indicate that collimation will indeed be achieved, but what of the observational situation?

The first optical Type I GRB afterglow, that of GRB 050709, was observed to very faint magnitudes (Hjorth \etal\ \cite{Hjorth2005}; Fox \etal\ \cite{Fox2005}), but no jet break was found in these observations (Watson \etal\ \cite{Watson2006}). As we detailed above, Type I GRBs exhibit very faint afterglows in the optical, and often, detections at late times are additionally hampered by the influence of the host galaxy (though some GRBs are offset so strongly that they do not lie on the host galaxy light, of course). Therefore, the best strategy to pursue the issue of Type I GRB collimation is to obtain late-time X-ray follow-up. While also less luminous than those of Type II GRBs in the X-rays, the corresponding reduction in X-ray-afterglow brightness is typically less extreme, and there are no issues of source confusion. Still, such observations are generally beyond the capabilities of the {\em Swift} satellite and therefore have to be performed by {\em Chandra} or {\em XMM-Newton}.

Two Type I GRBs with bright X-ray afterglows in 2005 were followed up at late times in such a way. GRB 050724 did not exhibit any break in its X-ray light curve until at least 2 Ms after the GRB, and Grupe \etal\ (\cite{Grupe2006}) derived an opening angle of $\gtrsim25^\circ$ from the observations, implying the afterglow was essentially uncollimated. On the other hand, Burrows \etal\ (\cite{Burrows2006}) report on the detection of a clear break in the light curve of the energetic Type I GRB 051221A, implying a jet opening angle of $\Theta_{\textnormal{jet}}\approx4-8^\circ$.

The single case of significant detection of steep late-time decay in the optical is the extremely luminous GRB 090510, first reported by McBreen \etal\ (\cite{McBreen2010}), and confirmed by the more thorough analysis of Nicuesa Guelbenzu \etal\ (\cite{Nicuesa2012A}). This decay has been interpreted to be post-break, but the theoretical interpretation of the complicated multi-wavelength afterglow yields inconclusive results (Kumar \& Barniol Duran \cite{Kumar2010}; De Pasquale \etal\ \cite{DePasquale2010}). If due to an actual jet break, an extreme collimation of $\approx1^\circ$ is implied.

Recently, further late-time observations of X-ray afterglows have been reported. Fong \etal\ (\cite{Fong2012}) studied the afterglow of GRB 111020A, deriving the existence of a break from a late, deep {\em Chandra} non-detection, and computing an opening angle similar to that of GRB 051221A. Another counterexample was found in GRB 120804A (Berger \etal\ \cite{Berger2012}; Troja \etal\, in preparation), which shows a non-breaking afterglow to out beyond 4 Ms. We show the light curves of all four GRB afterglows in Fig. \ref{Xray}.

Nicuesa Guelbenzu \etal\ (\cite{Nicuesa2012B}) compared the jet-opening-angle distribution of Type II and Type I GRBs and concluded that while there is an indication for a wider distribution for Type I GRBs, the issue is still hampered by the unknown distribution of the circumburst medium density, the lack of redshifts, etc. More precise values could be derived with the help of radio observations, but Type I GRBs are almost never detected in the radio, as very deep limits on two above-mentioned GRBs show (Fong \etal\ \cite{Fong2012}; Berger \etal\ \cite{Berger2012}).

\section{The existence of {\em dark} Type I GRB afterglows}

An afterglow is called ``dark'' when the optical luminosity is suppressed with respect to a conservative (usually $\beta_X-0.5$) extrapolation of the X-ray luminosity into the optical range (Jakobsson \etal\ \cite{Jakobsson2004}; Role \etal\ \cite{Rol2005}; van der Horst \etal\ \cite{vanderHorst2009}), under the assumption that the external (forward) shock is responsible for the afterglow emission (e.g., Sari \etal\ \cite{Sari1998}). This is often congruent with an optical non-detection despite deep and fast follow-up, but even optically bright afterglows can be dark according to the given definition, an example is the highly extinguished but ultra-luminous afterglow of GRB 080607 (Perley \etal\ \cite{Perley2011}). In the case of Type II GRBs, extensive studies have revealed most dark GRBs are due to rest-frame extinction in the GRB host galaxy, either local and patchy, or globally in highly reddened galaxies (e.g., Perley \etal\ \cite{Perley2009}; Greiner \etal\ \cite{Greiner2011}; Kr\"uhler \etal\ \cite{Kruehler2011}; Rossi \etal\ \cite{Rossi2012}), only a small number occur at very high redshifts and are dark due to Lyman absorption being redshifted into the optical.

One would naively expect darkness not to be an issue for Type I GRB afterglows; after all, they should usually occur far from the birthplaces of the massive stars that created the compact objects which represent the progenitor system of the GRBs. One of the assumptions Kann \etal\ (\cite{Kann2011}) made in the cases where afterglow data did not allow the rest-frame extinction to be constrained (i.e., almost all cases) was that $A_V=0$ mag. But this need not always to be the case, and evidence is mounting that some Type I GRB afterglows are reddened, or even truly dark.

One of the first indications was found by Ferrero \etal\ (\cite{Ferrero2007}) when studying the SED of GRB 050709, it exhibited a steep spectral slope in the optical and a strong curvature when combined with an NIR detection\footnote{While this detection was of low significance, it was significantly deeper than expected if the optical slope were just extrapolated into the NIR assuming a significantly lower extinction value. Since no SED including the X-rays was constructable, the result is to be taken with caution, though.}, implying a large rest-frame extinction $A_V\approx0.7$ mag, and this despite the large offset from its host galaxy (Fox \etal\ \cite{Fox2005}). The first clear evidence for a Type I GRB afterglow affected by host-galaxy dust extinction was the very red afterglow of GRB 070724A, discovered by Berger \etal\ (\cite{Berger2009}), which exhibits $A_V\approx0.9-1.3$ mag.	Kann \etal\ (\cite{Kann2011}) also find evidence for even higher extinction ($A_V\approx1.5$ mag) in the case of GRB 070809. Such a value is large even compared to most Type II GRB afterglows with definite extinction measurements (e.g., Kann \etal\ \cite{Kann2010}, Kr\"uhler \etal\ \cite{Kruehler2011}).

More recently, several Type I GRBs with deep observations all exhibit evidence for significant rest-frame extinction. Both GRB 111020A (Fong \etal\ \cite{Fong2012}) and GRB 120804A (Berger \etal\ \cite{Berger2012}) have already been mentioned in the context of the jet-break question. GRB 111020A shows evidence for a very high neutral-hydrogen column density from X-ray observations, though a direct link to the optical extinction cannot be made, as the dust-to-gas ratio in the host galaxy is unknown. In the case of GRB 120804A, an optical afterglow is discovered, and is found to be strongly suppressed vs. the X-ray afterglow, making the GRB classically dark, the required extinction is very high, $A_V\approx2.5$ mag. Additionally, this GRB is extraordinary as it is the first GRB that has been detected in an Ultra-Luminous InfraRed Galaxy\footnote{The Type I GRB 100206A had already been found in a less-Luminous InfraRed Galaxy (Perley \etal\ \cite{Perley2012}).}. A third recent example is GRB 111117A (Margutti \etal\ \cite{Margutti2012}; Sakamoto \etal\ \cite{Sakamoto2012}), which also exhibits evidence for a high neutral-hydrogen column density from X-ray observations, and $A_V\gtrsim0.5$ mag is implied. In all these cases, the evidence for a dense sightline clashes with the host-galaxy offset derived from the subarcsecond optical or {\em Chandra} X-ray positions\footnote{We note that it could be possible that these GRBs originate in very faint background galaxies, wherein they are more deeply embedded. In that case, though, the redshift is very likely to be even higher, increasing the derived neutral hydrogen column density as well as the optical extinction. Furthermore, it seems unlikely that such a projection effect would affect all the cases mentioned above.}. Deeper observations will be needed to elucidate why the afterglows of Type I GRBs are resembling those of Type II GRBs more and more.

%%-----------------------------
%%      your bibliography
%%-----------------------------

\end{document}